\documentclass[english]{article}
\usepackage{times}
\usepackage[T1]{fontenc}
\usepackage[latin1]{inputenc}
\usepackage{a4wide}
\usepackage{amsmath}
\usepackage{graphicx}

\makeatletter

\usepackage{babel}
\makeatother
\begin{document}

\title{Sampling errors of correlograms with and without sample mean removal
for higher-order complex white noise with arbitrary mean}

\author{T. D. Carozzi and A. M. Buckley}

\author{Space Science Centre, Sussex University, Brighton BN1 9QT, United
Kingdom}

\date{5 June 2005}

\maketitle
\begin{abstract}
We derive the bias, variance, covariance, and mean square error of
the standard lag windowed correlogram estimator both with and without
sample mean removal for complex white noise with an arbitrary mean.
We find that the arbitrary mean introduces lag dependent covariance
between different lags of the correlogram estimates in spite of the
lack of covariance in white noise for non-zeros lags. We provide a
heuristic rule for when the sample mean should be, and when it should
not be, removed if the true mean is not known. The sampling properties
derived here are useful is assesing the general statistical performance
of autocovariance and autocorrelation estimators in different parameter
regimes. Alternatively, the sampling properties could be used as bounds
on the detection of a weak signal in general white noise.
\end{abstract}

\section{Introduction}

The correlogram, an estimate of the autocorrelation function (ACF)
of a time series, is one of the cornerstone analyses in the signal
processing toolbox and therefore an understanding of its statistical
errors for various random signals is of fundamental importance. Since
the first published work \cite{Bartlett46} the sampling properties
of various ACF estimators for a wide variety of different processes
have been investigated \cite{Marriott54,Jenkins68,Anderson71}. Yet
it seems that one fundamental process has not draw full attention:
general independent and identically distributed (IID) processes. In
particular, IID processes introduce two novel aspects to the statistical
errors of ACF estimators: a non-zero mean and non-analytic signals.

In order to investigate these novel effects and in particular the
non-zero mean in particular, we compare the standard correlogram estimator
both with and without sample mean subtracted from the data samples.
These two estimators are introduced in section \ref{sec:Classic-Autocorrelation-Estimators}
and section \ref{sec:Classic-Autocovariance-Estimators} respectively.
We conclude by discussing how to treat the mean of a process when
estimating correlation.

Our main finding is that an unknown mean introduces non-zero lag dependent
covariance irrespective of whether the sample mean is removed or not.
This is surprising seeing that white noise signals are not non-zero
dependent.

\section{Lag windowed correlogram estimators}

In dealing with an uncertain mean when estimating the autocovariance
of a give sequence of data there are two possible procedures: either
one estimates the mean from the given data or the mean can be guessed
in some way not based on the given data. These two procedures taken
together with the subsequent correlogram computation can be seen as
two different types of autocovariance estimators. That is, there are
autocovariance estimators that include some sort of mean estimation
and there correlograms that do not. In what follows we consider one
version of each, namely, we consider the standard windowed correlogram
and the standard windowed correlogram with sample mean removal.

\subsection{Estimator without sample mean removal \label{sec:Classic-Autocorrelation-Estimators}}

The classical correlogram does not involve any explicit mean estimation.
Allowing for lag windowing, we define it for positive lags as

\begin{equation}
\widehat{R}_{zz}^{(\mu)}[l]:=\mathcal{W}[l]\sum_{k=1}^{N-l}z^{*}[k]z[k+l],\quad l=0,1,2...N-1,\label{eq: ACS_classicalEst}\end{equation}
where $z[1],z[2],...z[N]$ are possibly complex data samples, $N$
is the number of data samples, $\mathcal{W}[l]$ is an arbitrary real
valued lag weighting function, and $z^{*}$ denotes the complex conjugate
of $z$. We use square brackets $[\cdot]$ to refer to a function
of a discrete valued variable.

There are two standard choices for $\mathcal{W}[l]$. If\begin{equation}
\mathcal{W}[l]=\mathcal{W}_{U}[l]:=\frac{1}{N-|l|}\label{eq:UnbiasedWeighting}\end{equation}
 the estimators is called the traditional unbiased correlogram, while
\begin{equation}
\mathcal{W}[l]=\mathcal{W}_{P}[l]:=\frac{1}{N}\label{eq:PeriodogramWeighting}\end{equation}
 is known as the asymptotically unbiased correlogram. The latter is
exactly the Fourier transform of the periodogram. The negative lags
of the correlogram are determined from the positive lags in (\ref{eq: ACS_classicalEst})
through\begin{equation}
\widehat{R}_{zz}^{(\mu)}[l]:=\left(\widehat{R}_{zz}^{(\mu)}[-l]\right)^{*},\quad l=-1,-2,...,-N+1.\label{eq: negLagByConj_ACS}\end{equation}

The correlogram $\widehat{R}_{zz}^{(\mu)}$ can be interpreted as
an estimator of either the autocovariance sequence (ACVS) or the autocorrelation
sequence (ACS) of the discrete-time, complex-valued random variable
sequence $\{ Z[n];\: n=...-2,-1,0,1,2...\}$. By the ACS of $\{ Z[l]\}$,
we mean explicitly the function\[
R_{ZZ}[l]:=\mathrm{E}\left\{ Z^{*}[n]Z[n+l]\right\} \]
where $\mathrm{E}\left\{ \cdot\right\} $ represents the expectation
operator and $Z^{*}$ denotes the complex conjugate of $Z$; and by
the ACVS of $\{ Z[n]\}$, we mean\[
C_{ZZ}[l]:=\mathrm{Cov}\left\{ (Z[n]-\mu),(Z[n+l]-\mu)\right\} =\mathrm{E}\left\{ (Z[n]-\mu)^{*}(Z[n+l]-\mu)\right\} \]
see \cite{Bendat00} for details. Thus, the ACVS differs from ACS
in that the mean $\mu$ has been removed from the data sequence. In
other words, ACVS is equivalent to its ACS if the mean of the data
sequence is zero. For continuously sampled processes, the ACS and
ACVS are known as the autocorrelation function (ACF) and autocovariance
function (ACVF) respectively.

Note that $\widehat{R}_{zz}^{(\mu)}$, for a nonzero mean $\mu$ signal,
can also be interpreted as an autocovariance estimate in which, based
on other, separate information, an assumed mean $\mu_{g}$ was subtracted
from the data (or that $\mu_{g}$ was assumed 0 and nothing was done
to the data) which actually had the mean $\mu_{t}$. In other words,
$\mu=\mu_{t}-\mu_{g}$ could be seen as the error in the assumed mean.
If the mean $\mu$ of the sequence $\{ z[n]\}$ is known exactly the
signal can be converted into a zero mean sequence $\{ z'[n]\}:=\{ z[n]-\mu\}$
by subtracting the mean. We distinguish this case of the estimator
$\widehat{R}_{zz}^{(\mu)}[l]$ by replacing the $(\mu)$ index with
$(0)$, viz $\widehat{R}_{z'z'}^{(0)}[l]$. Thus we can summarize
formally the relationship between the autocorrelation and autocovariance
estimators mentioned above as

\[
\widehat{R}_{z'z'}^{(0)}[l]=\widehat{C}_{zz}[l]\]
which says that the estimator of the form (\ref{eq: ACS_classicalEst})
is an ACVS estimator if the process has a zero mean. The sampling
properties of this special case estimator, $\widehat{R}_{zz}^{(0)}[l]$,
is well documented and goes back to \cite{Bartlett46}. The sampling
properties of $\widehat{R}_{zz}^{(\mu)}$ with nonzero $\mu$ however,
is what will concern us subsequently.

\subsection{Estimator with sample mean removal \label{sec:Classic-Autocovariance-Estimators}}

If we extend the standard correlogram introduced in the previous section
to include the subtraction of the sample mean from data samples we
get the ACVS estimator

\begin{equation}
\widehat{C}_{zz}^{(\overline{\mu})}[l]:=\mathcal{W}[l]\sum_{k=1}^{N-l}(z[k]-\overline{z})^{*}(z[k+l]-\overline{z}),\quad l=0,1,2...N-1,\label{eq: ACVS_classicEst}\end{equation}
where\[
\overline{z}:=\frac{1}{N}\sum_{k=1}^{N}z[k]\]
is the sample mean. The negative lags can be estimated through a formula
analogous to (\ref{eq: negLagByConj_ACS}).

There are alternatives to simply subtracting the sample mean as in
(\ref{eq: ACVS_classicEst}) when estimating the ACVS if the mean
is not known, see e.g. \cite{Marriott54}. Here we will only consider
$\widehat{C}_{zz}^{(\overline{\mu})}$ as it is still commonly used
and of fundamental importance.

\section{General complex higher-order white noise}

The correlograms will now be applied to higher-order complex white
noise with arbitrary mean. We seek to derived the sampling properties
of each correlogram up to second-order properties. As it turns out,
we only need to consider moments of the process up to fourth order.
Thus, for our purposes it suffices to define a test sequence $\epsilon[\cdot]$
with the following properties

\begin{align}
E\{\epsilon[n]\} & =\mu\\
\mathrm{Cov}\left\{ \epsilon'[n],\epsilon'[n+l]\right\} =\mathrm{Cov}\left\{ \epsilon'[n]^{*},\epsilon'[n+l]^{*}\right\}  & =\sigma^{2}\delta_{0,l}=C_{\epsilon\epsilon}[l]\label{eq:ACVSwhitenoise}\\
\mathrm{Cov}\left\{ \epsilon'[n]^{*},\epsilon'[n+l]\right\}  & =m_{2}\delta_{0,l}=s^{2}\exp(i\theta_{2})\delta_{0,l}\\
E\left\{ \epsilon'[n]^{*}\epsilon'[n+l]\epsilon'[n+l']\right\}  & =\kappa_{3}\delta_{0,l}\delta_{0,l'}\\
\mathrm{Cov}\left\{ \epsilon'[n]^{*}\epsilon'[n+l],\epsilon'[n']^{*}\epsilon'[n'+l']\right\}  & =\kappa_{4}\delta_{n,n'}\delta_{0,l}\delta_{0,l'}+\sigma^{4}\delta_{n,n'}\delta_{l,l'}+s^{4}\delta_{n+l,n'}\delta_{-l,l'}\end{align}
where $\epsilon':=\epsilon-\mu$ is the centralised version of the
process, $\delta_{l,m}$ is the Kronecker delta, $\mu$ is the mean,
$\sigma^{2}$ is the central variance, $m_{2}$ is the second central
moment, $\kappa_{3}$ and $\kappa_{4}$ are the third and fourth order
cumulants respectively and $n$, $n'$, $l$ and $l'$ are all arbitrary
integers. The quantity $m_{2}$ and $s^{2}$ is what we will call
the \emph{quadratic variance} and the \emph{quadratic variance amplitude}
respectively. These names reflect the fact that they are not hermitian
in contrast with the ordinary variance $\sigma^{2}$.

The first property implies that the process has an arbitrary mean.
The second is that the autocovariance is zero except for the zero
lag. The third property is the nonhermitian quadratic autocovariance
of the process which usually is either zero or equal to the autocovariance.
The fourth property is a third order lagged cumulant of the process.
The last property basically implies that there is no covariance between
different autocovariance lags. Fifth orders and above are not specified
and so $\epsilon[\cdot]$ could have higher order correlation. Thus
$\epsilon[\cdot]$ is more general than IID processes yet is equivalent
in fourth and lower order properties. See \cite{Nikias93} for definitions
of higher order lagged cumulants of random processes.

For reference, we have the following relations in special cases: for
a purely real process (hence nonanalytic),\[
\Im\{\epsilon[n]\}=0\;\Rightarrow\; s^{2}=\sigma^{2}\]
where $\Im\{\cdot\}$ is the real part operator, while for an analytical
process \cite[sec. 13.2.3]{Bendat00}\[
\epsilon[\cdot]\,\mathrm{is}\,\mathrm{analytic}\;\Rightarrow\; s=0\]
 and for zero mean complex circular Gaussian white noise\[
\forall k,\,\epsilon[k]\sim\mathcal{N}\{0,\sigma\}\;\Rightarrow\;\kappa_{4}=\kappa_{3}=s=\mu=0,\]
i.e. only $\sigma^{2}$is non-zero, while for nontrivial Poissonian
real white noise\[
\forall k,\,\epsilon[k]\sim\mathrm{Poi}\{\lambda\}\;\Rightarrow\;\kappa_{4}=\kappa_{3}=\sigma^{2}=\mu=\lambda,\, s=0.\]

\section{Sampling properties of the estimators to second order}

We now present some of the sampling properties, namely the bias, the
variance, the covariance, and the mean-square error (MSE) for the
case of the general noise process $\epsilon[\cdot]$. These quantities
were derived employing the usual techniques of estimation theory \cite{Bendat00}
and assuming that the process has the properties given in the previous
section. The details of the derivations are given in a companion paper
\cite{Carozzi05a}.

\subsection{\label{sub:PropCorrNoSampMean}Sampling properties of correlogram
without sample mean removal}

The estimator $\widehat{R}^{(\mu)}$ was defined in (\ref{eq: ACS_classicalEst}).
With respect to the $\epsilon[\cdot]$ process, its sampling properties
to second order are as follows. The bias for all lags is found to
be

\begin{equation}
\mathrm{Bias}\left\{ \widehat{R}_{\epsilon\epsilon}^{(\mu)}[l]\right\} =E\left\{ \widehat{R}_{\epsilon\epsilon}^{(\mu)}[l]\right\} -R_{\epsilon\epsilon}[l]=((N-|l|)\mathcal{W}[l]-1)(\sigma^{2}\delta_{0l}+|\mu|^{2})\end{equation}
This shows that the so called unbiased estimator, $\mathcal{W}_{U}[l]$,
is in fact always unbiased even when the mean is not zero. All other
$\mathcal{W}[l]$ will result biased estimators if the mean is not
zero.

The covariance/variance of $\widehat{R}^{(\mu)}$ for any two lags
$l$ and $l'$ of the same sign is\begin{align}
\mathrm{Cov}\left\{ \widehat{R}_{\epsilon\epsilon}^{(\mu)}[l],\widehat{R}_{\epsilon\epsilon}^{(\mu)}[l']\right\}  & =\mathcal{W}[l]\mathcal{W}[l']\left(\delta_{0l}\delta_{0l'}(\kappa_{4}+s^{4})N+2\delta_{0l}\Re\{\mu\kappa_{3}^{*}\}(N-|l'|)+\right.\nonumber \\
 & \quad+2\delta_{0l'}\Re\{\mu\kappa_{3}^{*}\}(N-|l|)+\delta_{ll'}\sigma^{4}(N-|l|)+2|\mu|^{2}\sigma^{2}\left(N-\max(|l|,|l'|)\right)\nonumber \\
 & \quad\left.+2\Re\{\mu^{2}m_{2}^{*}\}(N-\min(|l|+|l'|,N))\right),\qquad ll'\ge0\end{align}
and when the lags have different signs the covariance is\begin{align}
\mathrm{Cov}\left\{ \widehat{R}_{\epsilon\epsilon}^{(\mu)}[l],\widehat{R}_{\epsilon\epsilon}^{(\mu)}[l']\right\}  & =\mathcal{W}[l]\mathcal{W}[l']\left(\delta_{0l}\delta_{0l'}(\kappa_{4}+\sigma^{4})N+2\delta_{0l}\Re\{\mu\kappa_{3}^{*}\}(N-|l'|)+\right.\nonumber \\
 & \quad+2\delta_{0l'}\Re\{\mu\kappa_{3}^{*}\}(N-|l|)+\delta_{-l,l'}s^{4}(N-|l|)+2|\mu|^{2}\sigma^{2}\left(N-\max(|l|,|l'|)\right)\nonumber \\
 & \quad\left.+2\Re\{\mu^{2}m_{2}\}(N-\min(|l|+|l'|,N))\right),\qquad ll'\le0\end{align}
These expressions can broken down into all combinations of covariances
and variances for zero lags and nonzero lags. The terms without $\delta$
factors are the covariance of the nonzero lags of the estimator. They
are zero if the mean is zero but otherwise they are non-zero and have
a piece-wise linear dependence on the lags before applying the weight
functions $\mathcal{W}[l]\mathcal{W}[l']$. The $\delta_{ll'}$ term
plus the terms without $\delta$ for $l=l'\neq0$ constitute the variance
of the nonzero lags. This is nonzero and linear in lag before weighting
even when the mean is zero. The terms with single $\delta_{0l}$ or
$\delta_{0l'}$ factors plus the terms without $\delta$ evaluated
at either $l=0$ or $l'=0$, are the covariance between zero lag and
non-zero lags. These are nonzero if the odd order moments $\mu$ and
$\mu_{3}$ are nonzero. Finally, the terms with the $\delta_{0l}\delta_{0l'}$
factor plus all other terms evaluated at $l=l'=0$ is the variance
of zero lag. It depends on all the moments. An example of the covariance
structure is shown in figure \ref{fig:AcfAcvfComp}a), \ref{fig:AcfAcvfComp}b),
and \ref{fig:AcfAcvfComp}c).

The bias and the variance given above can be combined to give the
mean-square error for the zero lag and the nonzero lag respectively\begin{align}
\mathrm{MSE}\left\{ \widehat{R}_{\epsilon\epsilon}^{(\mu)}[0]\right\}  & =(\kappa_{4}+s^{4}+4\Re\{\mu\kappa_{3}^{*}\}+2\Re\{\mu^{2}m_{2}^{*}\}+|\mu|^{4}N+\sigma^{2}(\sigma^{2}+2|\mu|^{2})(N+1))N\mathcal{W}^{2}[0]+\nonumber \\
 & \quad-2N(\sigma^{2}+|\mu|^{2})^{2}\mathcal{W}[0]+(\sigma^{2}+|\mu|^{2})^{2}\\
\mathrm{MSE}\left\{ \widehat{R}_{\epsilon\epsilon}^{(\mu)}[l\neq0]\right\}  & =\left(\sigma^{2}(\sigma^{2}+2|\mu|^{2})\left(N-|l|\right)+2\Re\{\mu^{2}m_{2}^{*}\}(N-\min(2|l|,N))+|\mu|^{4}(N-|l|)^{2}\right)\mathcal{W}^{2}[l]+\nonumber \\
 & \quad-2|\mu|^{4}(N-|l|)\mathcal{W}[l]+|\mu|^{4}\end{align}
These expressions are exact for all sample sizes $N$. Asymptotically,
that is as $N\rightarrow\infty$, the MSE behave, assuming $\mu\neq0$,
as\begin{align}
\mathrm{MSE}\left\{ \widehat{R}_{\epsilon\epsilon}^{(\mu)}[0]\right\}  & =\left(|\mu|^{2}+\sigma^{2}\right)^{2}\left(N^{2}\mathcal{W}^{2}[0]-2N\mathcal{W}[0]+1\right),\quad N\gg1,\:\mu\neq0\\
\mathrm{MSE}\left\{ \widehat{R}_{\epsilon\epsilon}^{(\mu)}[l\neq0]\right\}  & =|\mu|^{4}\left((N-|l|)^{2}\mathcal{W}^{2}[l]-2(N-|l|)\mathcal{W}[l]+1\right),\quad N\gg1,\:\mu\neq0\end{align}
where we have kept only the leading terms in $N$ and $l$ for each
power of $\mathcal{W}[l]$.

From the asymptotic expression, we see that for the unbiased lag weights
$\mathcal{W}_{U}[l]$ the leading terms in the MSE are zero. The asymptotic
MSE for the nonzero lags in this case is\begin{equation}
\mathrm{MSE}\left\{ \widehat{R}_{\epsilon\epsilon}^{(\mu)}[l\neq0]\right\} =\frac{\sigma^{2}(\sigma^{2}+2|\mu|^{2})+2\Re\{\mu^{2}m_{2}^{*}\}\left(N-\min(2|l|,N)\right)}{N-|l|},\quad\mathcal{W}\equiv\mathcal{W}_{U},N\gg1.\end{equation}
While for the periodogram weighting $\mathcal{W}_{P}[l]$, the asymptotic
MSE is $|\mu|^{4}l^{2}/N^{2}$ which does not tend to zero for large
lags and so it is not a consistent estimator.

This is in contrast to the well known case of zero mean. If $\mu=0$
then the asymptotic MSE of the nonzero lags is instead $\sigma^{4}(N-|l|)\mathcal{W}^{2}[l]$
for all $N$ as expected. The MSE of the unbiased estimator in this
case tends asymptotically to $\sigma^{4}/(N-|l|)$ and so it does
not converge for large lags, while the MSE for the periodogram estimator
tends to $\sigma^{4}(1-|l|/N)/N$ which tends to zero for large lags.
It is for this reason that the periodogram weighting function $\mathcal{W}_{P}[l]$
is preferred instead of the unbiased weighting $\mathcal{W}_{U}[l]$.

For the special case of zero mean see also \cite{Bartlett46}.

\subsection{Sampling properties of correlogram with sample mean removal}

The sampling properties of the correlogram estimator$\widehat{C}^{(\overline{\mu})}$,
defined in (\ref{eq: ACVS_classicEst}), for the noise process $\epsilon[\cdot]$
were found to be as follows.

The bias is

\begin{equation}
\mathrm{Bias}\left\{ \widehat{C}_{\epsilon\epsilon}^{(\overline{\mu})}[l]\right\} =\sigma^{2}\left((N\mathcal{W}[0]-1)\delta_{0l}-\frac{(N-l)\mathcal{W}[l]}{N}\right).\end{equation}
From this expression we find, remarkably, that all nonzero lags are
biased irrespective of the choice of weights. For instance, the weights
$\mathcal{W}_{U}[l]$ known as unbiased in relation to the zero-mean
correlogram $\widehat{R}^{(0)}$, lead to a bias of $-\sigma^{2}/N$
for all lags. It is however possible to get an unbiased estimate for
the zero lag if one chooses $\mathcal{W}[0]=1/(N-1)$. For this choice,
the zero lag of $\widehat{C}^{(\overline{\mu})}$ is equivalent to
the usual sample variance. 

The covariance/variance of $\widehat{C}^{(\overline{\mu})}$ between
any two lags $l$ and $l'$ of the same sign is\begin{align}
\mathrm{Cov}\left\{ \widehat{C}_{\epsilon\epsilon}^{(\overline{\mu})}[l],\widehat{C}_{\epsilon\epsilon}^{(\overline{\mu})}[l']\right\}  & =\mathcal{W}[l]\mathcal{W}[l']\left(\delta_{0,l}\delta_{0,l'}(\kappa_{4}+s^{4})N+\right.\nonumber \\
 & \quad-\delta_{0,l}\kappa_{4}\left(1-\frac{|l'|}{N}\right)-\delta_{0,l'}\kappa_{4}\left(1-\frac{|l|}{N}\right)+\delta_{l,l'}\sigma^{4}(N-|l|)+\nonumber \\
 & \quad+\frac{\kappa_{4}}{N}\left(1-\frac{2\max(|l|,|l'|)+2\min(|l|+|l'|,N)-3(|l|+|l'|)}{N}-3\frac{ll'}{N^{2}}\right)+\nonumber \\
 & \quad-\sigma^{4}\left(1-\frac{2\max(|l|,|l'|)-|l|-|l'|}{N}-\frac{ll'}{N^{2}}\right)+\nonumber \\
 & \quad\left.-s^{4}\left(1-\frac{2\min(|l|+|l'|,N)-|l|-|l'|}{N}-\frac{ll'}{N^{2}}\right)\right),\qquad ll'\ge0\end{align}
and for lags of different signs\begin{align}
\mathrm{Cov}\left\{ \widehat{C}_{\epsilon\epsilon}^{(\overline{\mu})}[l],\widehat{C}_{\epsilon\epsilon}^{(\overline{\mu})}[l']\right\}  & =\mathcal{W}[l]\mathcal{W}^{*}[l']\left(\delta_{0,l}\delta_{0,l'}(\kappa_{4}+\sigma^{4})N+\right.\nonumber \\
 & \quad-\delta_{0,l}\kappa_{4}\left(1-\frac{|l'|}{N}\right)-\delta_{0,l'}\kappa_{4}\left(1-\frac{|l|}{N}\right)+\delta_{|l|,|l'|}s^{4}(N-|l|)+\nonumber \\
 & \quad+\frac{\kappa_{4}}{N}\left(1-\frac{2\max(|l|,|l'|)+2\min(|l|+|l'|,N)-3(|l|+|l'|)}{N}-3\frac{|ll'|}{N^{2}}\right)+\nonumber \\
 & \quad-s^{4}\left(1-\frac{2\max(|l|,|l'|)-|l|-|l'|}{N}-\frac{|ll'|}{N^{2}}\right)+\nonumber \\
 & \quad\left.-\sigma^{4}\left(1-\frac{2\min(|l|+|l'|,N)-|l|-|l'|}{N}-\frac{|ll'|}{N^{2}}\right)\right),\qquad ll'\le0.\end{align}
Again, this expression unifies all combinations of variances and covariances
between zero and nonzero lags; see the discussion in the previous
subsection. The main differences with that of the estimator without
mean removal is that in this case there is no dependence on the odd
order moments $\mu$ and $\kappa_{3}$, and further, that the lag
dependence before applying the weighting functions factor is quadratic
in lag. An example of the covariance matrix is shown in figure \ref{fig:AcfAcvfComp}d),
\ref{fig:AcfAcvfComp}e), and \ref{fig:AcfAcvfComp}f).

The MSE of the $\widehat{C}^{(\overline{\mu})}$ estimator can be
determined for the expressions for the bias and covariance given above.
The result is\begin{align}
\mathrm{MSE}\left\{ \widehat{C}_{\epsilon\epsilon}^{(\overline{\mu})}[0]\right\}  & =\left(\sigma^{4}(N-1)N+(\kappa_{4}+s^{4})N-s^{4}-2\kappa_{4}+\frac{\kappa_{4}}{N}\right)\mathcal{W}^{2}[0]-2(N-1)\sigma^{4}\mathcal{W}[0]+\sigma^{4}\\
\mathrm{MSE}\left\{ \widehat{C}_{\epsilon\epsilon}^{(\overline{\mu})}[l\neq0]\right\}  & =\left(\sigma^{4}(N-|l|)\left(1-2\frac{|l|}{N^{2}}\right)+\frac{\kappa_{4}}{N}\left(1-\frac{2|l|+2\min(2|l|,N)-6|l|}{N}-3\frac{l^{2}}{N^{2}}\right)+\right.\nonumber \\
 & \quad\left.-s^{4}\left(1-\frac{2\min(2|l|,N)-2|l|}{N}-\frac{l^{2}}{N^{2}}\right)\right)\mathcal{W}^{2}[l]\end{align}
The MSE in the asymptotic case is\begin{align}
\mathrm{MSE}\left\{ \widehat{C}_{\epsilon\epsilon}^{(\overline{\mu})}[0]\right\}  & =\sigma^{4}\left(N^{2}\mathcal{W}^{2}[0]-2N\mathcal{W}[0]+1\right),\quad N\gg1\\
\mathrm{MSE}\left\{ \widehat{C}_{\epsilon\epsilon}^{(\overline{\mu})}[l\neq0]\right\}  & =\sigma^{4}(N-|l|)\mathcal{W}^{2}[l],\quad N\gg1\end{align}
where we have kept only the leading terms in $N$ or $l$ for each
coefficient of $\mathcal{W}[l]$. For the weighting sequence $\mathcal{W}_{U}[l]$,
the MSE tends to zero as $\sigma^{4}/(N-|l|)$; while for $\mathcal{W}_{P}[l]$,
the MSE tends to zero as $\sigma^{4}(1-|l|/N)/N$. So both weight
sequences lead to consistent estimators. However, the asymptotic MSE
in both cases is different for different lags.

For the special case of real valued processes see \cite{Anderson71}.

\section{Comparison of the correlograms with and without sample mean removal}

\begin{figure}
\includegraphics[%
  width=1.0\textwidth,
  keepaspectratio]{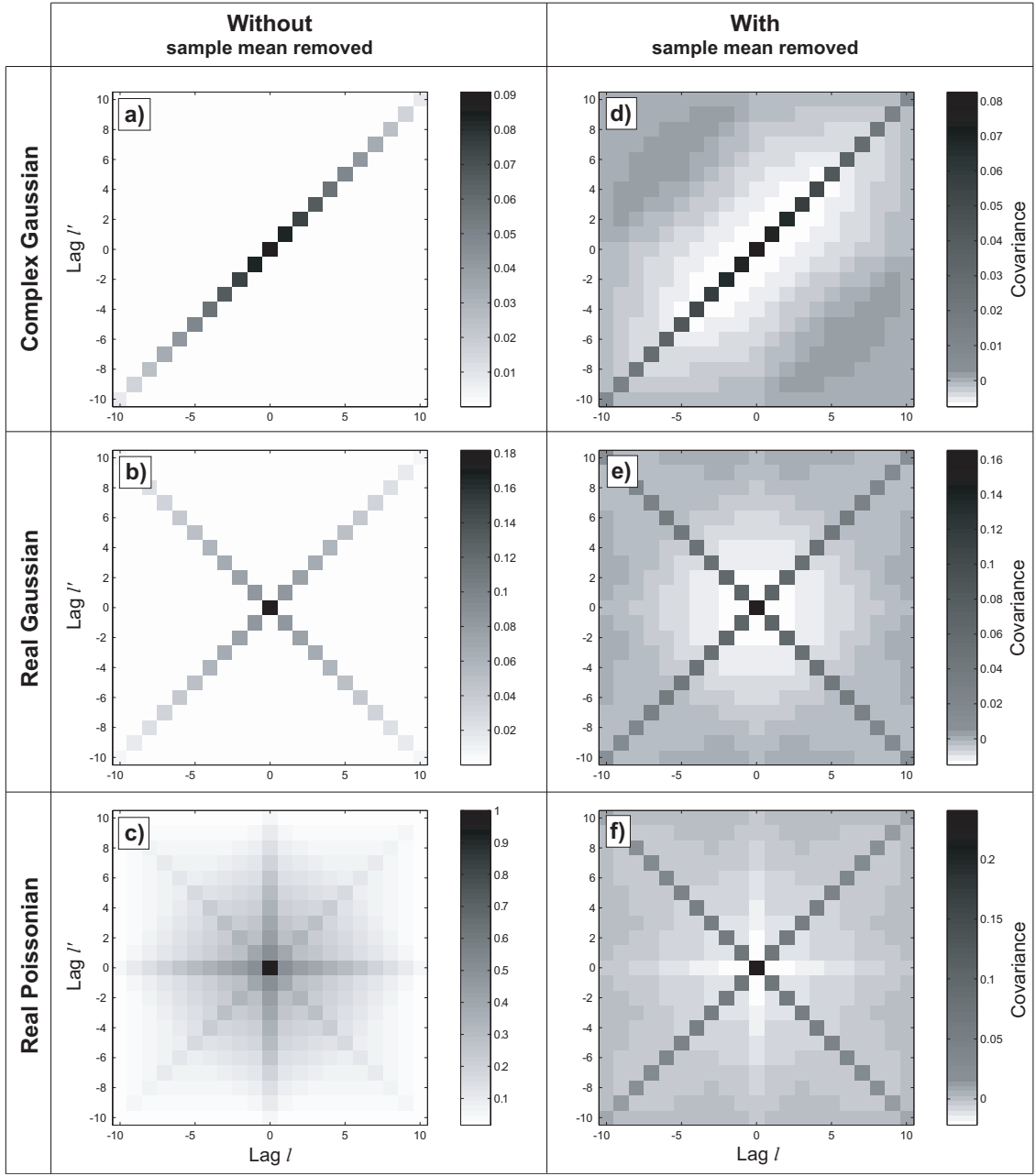}

\caption{\label{fig:AcfAcvfComp}Covariance and variance between any two lags
$l$ and $l'$ of the correlogram estimator ($N=11$) with and without
sample mean removal for various types of white noise signals. The
first column of panels, namely a), b) and c), are the correlogram
without sample mean removal, $\widehat{R}^{(\mu)}[l]$, defined in
equation (\ref{eq: ACS_classicalEst}); while the second column of
panels, namely d), e) and f), are the correlogram with sample mean
removal, $\widehat{C}^{(\overline{\mu})}[l]$, defined in equation
(\ref{eq: ACVS_classicEst}). The signal in the first row of panels,
namely a) and d) is analytic Gaussian white noise $\sigma^{2}=1$;
the second row, b) and e), is real Gaussian white noise; the third
row, c) and f), is real Poissonian white noise with $\mu=1$. In all
plots, the variance is along the $l=l'$ diagonal while the rest,
$l\neq l'$ is strictly covariance. Note that the gray-scale in the
second column is different from the linear scale of the first column
in order to enhance the structure of the covariance.}
\end{figure}

\begin{figure}
\includegraphics[%
  width=1.0\columnwidth]{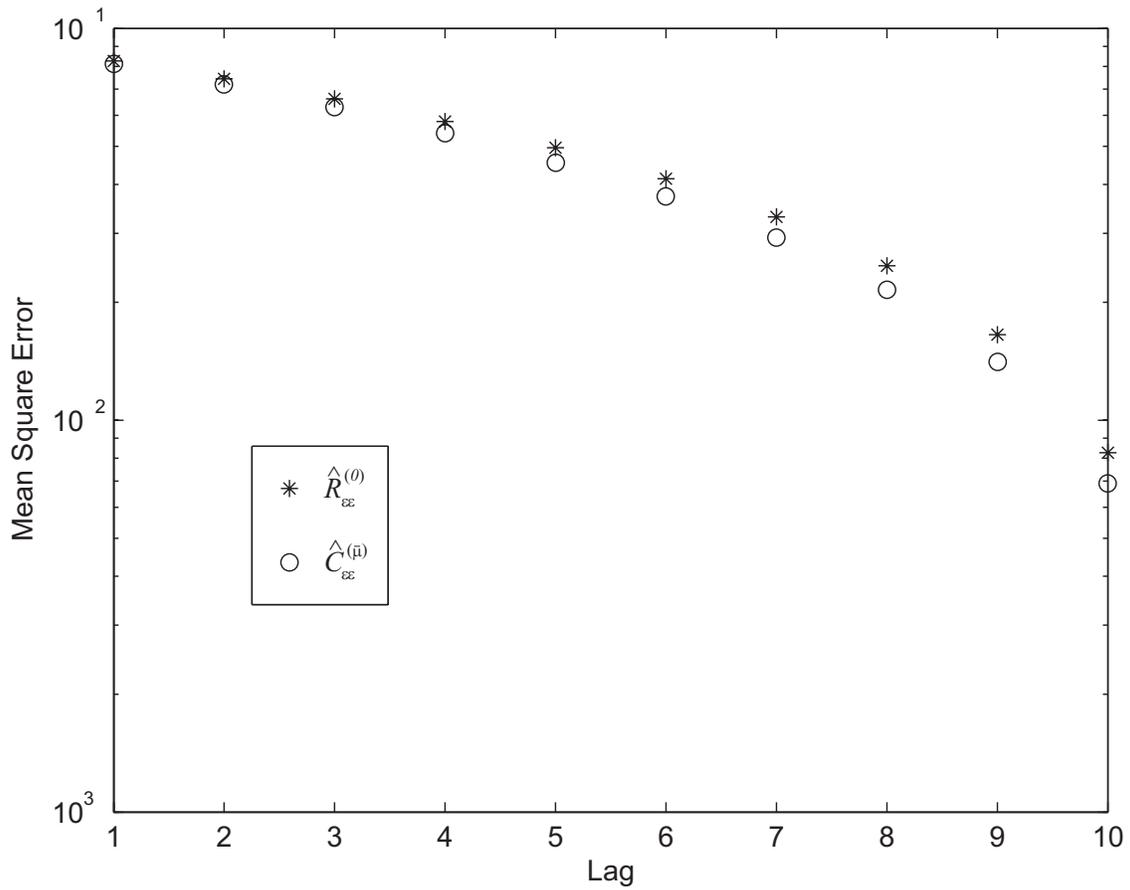}

\caption{\label{fig:ACFvsACVFmse}Comparison of the MSE of the $\widehat{R}^{(0)}$and
$\widehat{C}^{(\overline{\mu})}$estimators with $\mathcal{W}_{P}[l]$
weights for the case of analytic Gaussian white noise with $\sigma^{2}=1$.
For all lags, the autocovariance sequence estimator using the sample
mean, i.e. $\widehat{C}^{(\overline{\mu})}$represented by circles,
exhibits a smaller MSE than the estimator using the known mean, i.e.
$\widehat{R}^{(0)}$ represented by stars.}
\end{figure}

Now that we have the second order sampling of the correlograms with
and without sample mean removal we can compare them. Examples of the
covariance of the estimators are shown in figure \ref{fig:AcfAcvfComp}
for various kinds of white noise both with and without zero means;
and a comparison of the MSEs is shown in figure \ref{fig:ACFvsACVFmse}
for a zero mean noise process.

Inspection of the sampling properties derived here reveals some novel
features. One such feature is that the estimators nominally have nonzero
covariance, that is, the estimates at different lags are not independent.
This is down to the fact that the same data samples are reused in
the evaluation of different lags of the correlogram leading to a relationship
between lag estimates. More surprising is the fact that the variance
of both the estimators are nominally lag dependent. This is despite
the fact that the test sequence $\epsilon$ is white and hence has
no nonzero lag dependence, see (\ref{eq:ACVSwhitenoise}). Also the
covariance is lag dependent.

\subsection{$\widehat{R}^{(0)}$and $\widehat{C}^{(\overline{\mu})}$ as autocovariance
estimators when mean is known}

In the preceding sections we have assumed that the mean of the data
sequence we are trying to estimate the autocovariance of is unknown.
What can be said if the mean is known? It would seem natural that
if we knew the mean we would subtract it from the data samples and
use the estimator $\widehat{R}^{(0)}$. Thus we would not use the
estimator $\widehat{C}^{(\overline{\mu})}$ as it involves estimation
of the mean which we already know. Surprisingly however, inspection
of the sampling properties presented here show that this choice of
estimators is not immediately obvious.

The MSE of the estimator $\widehat{C}^{(\overline{\mu})}$ is, in
fact, smaller than that of the estimator $\widehat{R}^{(0)}$ for
general white noise, assuming the same lag window is used in both
estimators. The difference in the MSE is\begin{equation}
2\sigma^{4}\frac{l}{N^{2}}(N-l)\mathcal{W}^{2}[l]\end{equation}
 and is plotted in Figure \ref{fig:ACFvsACVFmse}. This difference
tends asymptotically to zero as $N$ is increased, but for finite
sample sizes the $\widehat{C}^{(\overline{\mu})}$ is always better
than the $\widehat{R}^{(0)}$ in the MSE sense. This is counter-intuitive
as it seems to violate the principle that the more one knows about
something, the better one can estimate it.

The solution to this conundrum is that although the error in the covariance
estimate is smaller, the error in the estimate of the mean is on the
other hand larger. Specifically the difference in the MSE in the mean
estimate is $\sigma^{4}/N$. If take, e.g., the weights $\mathcal{W}_{P}[l]$
and use the inequality $l<N$ then the MSE for lag $l$ must be smaller
than $2\sigma^{4}l/N^{3}$ and so the total MSE of the estimator is
of the order $\sigma^{4}/N$. This is comparable to the MSE in the
sample mean. Thus the total error, autocovariance and mean estimation,
is not better for the $\widehat{C}^{(\overline{\mu})}$ compared with
$\widehat{R}^{(0)}$. 

Furthermore $\widehat{C}^{(\overline{\mu})}$ has a nonzero covariance
between its lag estimates which $\widehat{R}^{(0)}$does not have.
This can be understood from the following observation: that the sum
over all non-negative lags of the estimator $\widehat{R}^{(\mu)}$
with periodogram weighting is equal to sample size times the square
of the sample mean, i.e.\[
{\displaystyle \sum_{l=0}^{N-1}}\widehat{R}_{zz}^{(P)}[l]=\frac{1}{N}{\displaystyle \sum_{l=0}^{N-1}}\sum_{k=1}^{N-l}z^{*}[k]z[k+l]=N\left|\sum_{k=1}^{N-1}z[k]\right|^{2}=N\left|\overline{z}\right|^{2}\]
for any process $\{ z[k]\}$. See \cite{Percival93b} for further
discussions. This shows that $\widehat{R}_{zz}^{(P)}$ and the sample
mean $\overline{z}$ are related. But since $\widehat{C}^{(\overline{\mu})}$
is based on $\widehat{R}_{zz}^{(P)}$ and $\overline{z}$ enters into
every lag estimate, this suggests that there is an interdependence
between lags and explain the nonzero covariance of the lags.

\subsection{When to use the sample mean if the mean is assumed small}

Let us now look at the situation when we believe that the mean is
small. The question is which estimator is better: the one without
or the one with sample mean removal. If we use the former we run the
risk that the error in the estimated mean is large than the true mean.
On the other hand, the mean may not be exactly zero so the latter
estimate will also be off. There is a trade off here and we wish to
find a criterion for when the sample mean should be removed.

Inspection of the results presented earlier suggest that as a rule
of thumb the two estimators are roughly equal when

\begin{equation}
|\mu|=\frac{\sigma}{\sqrt{2N}}\label{eq:RequaltoC_condition}\end{equation}
and when $|\mu|<\sigma/\sqrt{2N}$ , $\widehat{R}^{(\mu)}$ is preferable
to $\widehat{C}^{(\overline{\mu})}$ and when $|\mu|>\sigma/\sqrt{2N}$,
$\widehat{C}^{(\overline{\mu})}$ is preferable to $\widehat{R}^{(\mu)}$.
This condition can be understood in an intuitive way as follows. The
sample mean is an estimate of the population mean with a relative
error of $(\sigma/\sqrt{N})/|\mu|$. It makes sense to use this estimate
in the correlogram instead of the unknown mean only if this relative
error is smaller than 1, i.e. when $|\mu|>\sigma/\sqrt{2N}$. This
is because the absolute error if we do not remove anything is of course
only $|\mu|$. Thus we arrive at the condition as the found above
(\ref{eq:RequaltoC_condition}).

Naturally, if we do not know the mean we will not be able asses the
equality (\ref{eq:RequaltoC_condition}) exactly, but one could instead
get it approximately by using the sample mean and the sample variance
of the given data samples.

\section{Conclusion}

We have presented expressions for the bias, variance, covariance,
mean-square error of the classical correlogram estimator with and
without sample mean estimation for general complex white noise with
arbitrary mean. A summary of the sampling properties of the estimators
for $\epsilon[\cdot]$, i.e., complex higher order white noise with
arbitrary mean is as follows. The second-order sampling properties
of the correlogram without sample mean removal, $\widehat{R}^{(\mu)}$,
are in summary:

\begin{itemize}
\item $\epsilon[\cdot]$ moment dependence: cumulants up to fourth order
$\{\mu,\sigma^{2},s^{2},\kappa_{3},\kappa_{4}\}$ 
\item Bias: unbiased for $1/(N-l)$ weighting, even when $\mu\neq0$
\item Covariance: piecewise linear lag dependent covariance before weighting
\item Mean square error: is asymptotically proportional to $|\mu|^{4}$
when $\mu\neq0$ while if $\mu=0$ the MSE is asymptotically equal
to $\sigma^{4}(N-|l|)\mathcal{W}^{2}[l]$ 
\end{itemize}
For the correlogram with sample mean removal, $\widehat{C}^{(\overline{\mu})}$,
a summary of second-order sampling properties is:

\begin{itemize}
\item $\epsilon[\cdot]$ moment dependence: only even order cumulants $\{\sigma^{2},s^{2},\kappa_{4}\}$
\item Bias: nonzero lags biased for all weighting functions, zero lag unbiased
for $\mathcal{W}[0]=1/(N-1)$
\item Covariance: piecewise quadratic lag dependent covariance before weighting
\item Mean square error: asymptotically equal to $\sigma^{4}(N-|l|)\mathcal{W}^{2}[l]$
\end{itemize}
In terms of the properties of the process, we have found that a nonzero
mean can lead to covariance between lags of the correlogram, and that
complex valued processes lift the degeneracy between positive and
positive lags, (that is that the sampling properties of positive lags
and negative lags are different), and distinguishes between analytic
and non-analytic processes.

For both estimators, the both the variance and the covariance are
in general lag dependent. We conclude therefore that neither $\widehat{R}^{(\mu)}$
nor $\widehat{C}^{(\overline{\mu})}$ with the standard lag weights
$\mathcal{W}_{U}[l]$ and $\mathcal{W}_{P}[l]$ are optimal for estimating
the autocorrelation/autocovariance of noise processes with an unknown
mean.

Finally, we found that the sampling properties suggest that one should
always remove the mean if it is known apriori. If the mean is not
known, the sample mean should be removed only when $|\mu|>\sigma/\sqrt{2N}$,
i.e. $\widehat{R}_{\epsilon\epsilon}^{(\mu)}$ is preferable to $\widehat{C}_{\epsilon\epsilon}^{(\overline{\mu})}$
if $|\mu|<\sigma/\sqrt{2N}$.

\section*{Acknowledgments}

This work was sponsored by PPARC ref: PPA/G/S/1999/00466 and PPA/G/S/2000/00058.

\bibliographystyle{hplain}
\bibliography{main}

\end{document}